\documentclass{emulateapj}
\DeclareMathAlphabet{\mathsc}{OT1}{cmr}{m}{sc}

\begin{document}
\newcommand{\deep}{\textsc{deep2}}
\newcommand{\deimos}{\textsc{deimos}}
\newcommand{\cfht}{\textsc{cfht}}
\newcommand{\oii}{O\,\textsc{ii}}
\newcommand{\oiii}{O\,\textsc{iii}}
\newcommand{\nii}{N\,\textsc{ii}}
\newcommand{\sii}{S\,\textsc{ii}}
\newcommand{\lya}{Ly$\alpha$}
\newcommand{\ha}{H$\alpha$}
\newcommand{\hb}{H$\beta$}
\newcommand{\hg}{H$\gamma$}
\newcommand{\mathoii}{\mathrm{O\,\mathsc{ii}}}
\newcommand{\mathoiii}{\mathrm{O\,\mathsc{iii}}}
\newcommand{\mathha}{\mathrm{H}\alpha}
\newcommand{\mathhb}{\mathrm{H}\beta}

\title{The DEEP2 Galaxy Redshift Survey: \\ Redshift Identification of
Single-Line Emission Galaxies}

\author{Evan~N.~Kirby\altaffilmark{1}, Puragra~Guhathakurta,
  S.~M.~Faber, David~C.~Koo}
\affil{University of California Observatories/Lick Observatory,
  Department of Astronomy \& Astrophysics, \\
  University of California, Santa Cruz, CA 95064}
\author{Benjamin~J.~Weiner}
\affil{Steward Observatory, University of Arizona, Tucson, AZ 85721}
\and
\author{Michael~C.~Cooper}
\affil{Department of Astronomy, University of California, Berkeley, CA
  94720}

\altaffiltext{1}{\texttt{ekirby@ucolick.org}}

\shorttitle{\deep: Redshifts for Single-Line Galaxies}
\shortauthors{Kirby et al.}

\begin{abstract}
We present two methods for determining spectroscopic redshifts of
galaxies in the \deep\ survey which display only one identifiable
feature, an emission line, in the observed spectrum (``single-line
galaxies'').  First, we assume each single line is one of the four
brightest lines accessible to \deep: \ha, [\oiii] $\lambda 5007$, \hb,
or [\oii] $\lambda 3727$.  Then, we supplement spectral information
with $BRI$ photometry.  The first method, parameter space proximity
(PSP), calculates the distance of a single-line galaxy to galaxies of
known redshift in $(B-R)$, $(R-I)$, $R$, $\lambda_{\rm{observed}}$
parameter space.  The second method is an artificial neural network
(ANN).  Prior information, such as allowable line widths and ratios,
rules out one or more of the four lines for some galaxies in both
methods.  Based on analyses of evaluation sets, both methods are
nearly perfect at identifying blended [\oii] doublets.  Of the lines
identified as \ha\ in the PSP and ANN methods, 91.4\% and 94.2\%
respectively are accurate.  Although the methods are not this accurate
at discriminating between [\oiii] and \hb, they can identify a single
line as one of the two, and the ANN method in particular unambiguously
identifies many [\oiii] lines.  From a sample of 640 single-line
spectra, the methods determine the identities of 401 (62.7\%) and 472
(73.8\%) single lines, respectively, at accuracies similar to those
found in the evaluation sets.
\end{abstract}


\section{Introduction}
Photometric redshifts (photo-$z$s) save astronomers from expensive
spectroscopy by determining redshifts from efficient broadband
photometry.  However, photo-$z$ precision cannot compare to
spectroscopic redshift precision.  \citet{way06} compare five
state-of-the-art methods to determine redshifts from Sloan Digital Sky
Survey (SDSS) photometry.  Neural networks, which are non-linear
regression tools, perform the best, but the rms error in the
photo-$z$s is $\delta z/(1+z) \sim 0.02$ at best.  This precision is
sufficient to study large scale structure and some forms of redshift
evolution, but not local environments \citep{coo05}, kinematic pairs,
or the low-redshift luminosity function, where precise luminosities
require precise redshifts.  At a spectral resolution $R = 5000$, the
spectroscopic redshifts in the Deep Extragalactic Evolutionary Probe
(\deep) Galaxy Redshift Survey \citep{dav03} are many times more
precise.  Repeat observations show redshift errors of $\delta z \sim
10^{-4}$, or velocity errors of $\delta v \sim 30~\rm{km/s}$.

The expense of time-intensive spectroscopy demands a high rate of
successfully determined redshifts.  Nonetheless, some spectra fail at
providing redshifts.  One of the most common redshift failures is the
presence of only one emission line.  Ordinarily, recognizable patterns
of emission or absorption lines uniquely determine the lines'
identities and hence rest wavelengths.  Spectroscopically precise
observed wavelengths then give highly precise redshifts.  An isolated
emission line forms no recognizable pattern to reveal its identity and
rest wavelength.

Recovering failed spectroscopic redshifts with broadband photometry
has hardly been explored in the literature.  \citet{coh99} assume all
single emission lines in the Caltech Faint Galaxy Redshift Survey
(CFGRS) are [\oii] $\lambda 3727$ by arguing that any other line
except \lya\ would be accompanied by other emission lines.
\citet{lil95} have identified single emission lines in Canada-France
Redshift Survey (CFRS) spectra based on the slope of the continuum
surrounding the line.  (Unlike \deep, both CFGRS and CFRS did not have
the spectroscopic resolution to resolve the [\oii] doublet, requiring
them to make assumptions about the identities of single lines.)

In this paper, we also rely on the shape of the continuum, but we will
show that broadband $BRI$ colors can determine the identity of single
emission lines accurately, even for galaxies with no visible spectral
continuum.  The problem of an isolated emission line afflicts $\sim
1.5\%$ of \deep\ targets, but the problem is more prevalent for
serendipitously detected galaxies (``serendips'') that share a slit
with target galaxies through a fortuitous position on the sky.
Recovery of redshifts from serendip spectra is important because
serendips form an unbiased sample of galaxy spectra.

We explore the problem through two methods.  The parameter space
proximity (PSP) method identifies the redshift of a single-line galaxy
by comparing its photometric magnitude and colors with those of
galaxies with known spectroscopic redshifts.  The artificial neural
network (ANN) method employs the ANN machine learning algorithm, which
learns the functional relationship between any number of dependent and
independent variables.  They can determine photometric redshifts from
relevant observable quantities such as colors, apparent magnitudes,
and angular sizes.  Here, we use apparent magnitudes (from which the
ANN determines colors) and observed wavelengths of the single emission
lines.

This paper is organized as follows: \S\,\ref{sec:data} describes the
\deep\ survey and how single emission lines arise in its spectra;
\S\,\ref{sec:methods} describes the two methods for determining the
identities of these single lines; \S\,\ref{sec:accuracy} examines the
accuracy of each method; \S\,\ref{sec:results} presents the results of
applying the methods to single emission lines; and
\S\,\ref{sec:conclusions} summarizes our work and discusses our future
plans.

\section{Data}
\label{sec:data}
\subsection{DEEP2 Survey}
The \deep\ Survey \citep[outlined by][]{dav03} combines $BRI$
photometry in four fields \citep[described by][]{coi04} from the
\cfht\ $12\mathrm{k} \times 8\mathrm{k}$ mosaic camera and
spectroscopy (described by Newman et al.\ 2007, in preparation) from
the \deimos\ spectrograph \citep{fab03} on the Keck {\sc ii}
telescope.  \citet{dav05} detail the \deep\ target selection, based on
the three-filter ($B$, $R$, $I$) photometry.  A color-color cut is
applied to spectroscopy candidates in three of the four 120
$\mathrm{arcmin}^2$ \deep\ fields to reduce as much as possible the
number of spectra of $z < 0.7$ galaxies.  The remaining field (Field
1) has no such color-color selection.  An apparent magnitude cut of
$R_{\mathrm{AB}} < 24.1$ is applied to all fields.  Furthermore, each
spectroscopy target must be detected in $B$, $R$, and $I$.  The
\texttt{spec2d} and \texttt{spec1d} software packages, written by the
\deep\ team, accomplish the \deimos\ spectroscopy reduction, including
sky subtraction and an instrumental throughput correction to preserve
actual line strengths.

Field 1 is unique in other ways as well.  It overlaps the Extended
Groth Strip (EGS), which many different teams and instruments observe
heavily.  The wealth of photometry for the \deep\ targets in Field 1
permits very good photo-$z$ estimates (Huang et al.\ 2007, in
preparation).  We plan to make use of the additional EGS data in
future work.  However, in this paper, we use only the $BRI$ photometry
in Field 1, and we treat it no differently from Fields 2-4.

Astronomers from the \deep\ team visually inspect every spectrum using
the \texttt{zspec} script from the \texttt{spec1d} software package,
designed by \deep\ members.  They assign spectra with at least two
strong identifiable lines---such as an unblended [\oii] doublet---a
redshift quality code $Q = 4$.  Spectra with one strong line and at
least one weaker line receive $Q = 3$.  The entire survey contains
49,059 spectra.  Of them, 27,460 and 5,829 are $Q = 4$ and $3$
respectively.  The $Q = 2$ category encompasses all failed redshifts
that may be recovered with additional effort.  The inspectors note the
reason for all failures in this category, including the absence of all
but one line.

\subsection{Single Emission Line Galaxies}
\label{sec:sngls}
The spectral resolution of a \deep\ spectrum is 1.4~\AA\ FWHM, and the
typical spectral range is about 6600~\AA\ to 9200~\AA.  Therefore, few
redshifts permit both \ha\ $\lambda 6563$ and [\oiii] $\lambda 5007$
or \hb\ $\lambda 4861$ to fall on the same spectrum.  Similarly, few
redshifts permit both [\oiii] or \hb\ and [\oii] $\lambda 3727$.
Consequently, most \deep\ galaxies which are too faint to have a
weaker emission line or a continuum with noticeable absorption lines
display only one visible feature: \ha, [\oiii], \hb, or [\oii] in
emission.  This feature is often the [\oii] doublet, which is easily
identifiable by its invariable $220~\mathrm{km/s}$ peak separation as
long as the galaxy's internal gas velocity dispersion does not broaden
and blend both peaks.  However, the feature may be a truly isolated
single line, in which case the spectrum cannot yield a unique
redshift.  Lines may appear to be orphans if the signal-to-noise is
low enough to permit only the brightest line to be seen.  Other weak
lines may be lost in the noise of night sky lines, even with good sky
subtraction.  Finally, a gap of $\sim 5$~\AA\ separates the red and
blue CCDs in \deimos, and any line in a pair of otherwise visible
lines that falls completely in the gap will orphan its partner.

We select all 984 redshift failures that result from a single emission
line.  After an additional visual inspection to remove spectra with
marginally detected lines, spectra mistakenly marked as having a
single line, and spectra without any visible emission lines, our
sample contains 640 single-line emission galaxies.  We identify the
pixel within the line that contains the most counts in the spectrum
smoothed through a Gaussian window function with $\sigma =
4~\mathrm{pixels} = 1.3$~\AA\ and inverse variance weighting.  We
designate that pixel's observed wavelength as the wavelength of the
single line.  Finally, we note whether the line is broad enough to be
a blended [\oii] doublet.  For the purposes of identifying lines, it
is not necessary to be more precise in determining the observed
wavelength of the [\oii] doublet than selecting the pixel with the
most counts.

It is worth mentioning that the line identification methods presented
here will fail at identifying the redshifts of composite galaxies,
such as those blended together through lensing or line-of-sight
coincidence.  We assume that the broadband colors and magnitudes
associated with all of the lines identified come from a single galaxy,
and that composite colors and magnitudes will be distinct enough in
parameter space that their associated single lines will not receive a
conclusive identification.

\section{Methods}
\label{sec:methods}
The bright lines most often visible in \deep\ spectra are \ha\
$\lambda 6563$, [\oiii] $\lambda 5007$, \hb\ $\lambda 4861$, and the
[\oii] $\lambda\lambda 3726$, 3729 doublet.  If any other line is
visible, then one of these four is almost always visible as well.  The
$B$-band detection requirement eliminates the possibility of \lya.  On
rare occasions, bright sky lines or the \deimos\ CCD gap may hide one
of these four lines, leaving only one other dimmer line, such as
[\oiii] $\lambda 4959$ or \hg\ $\lambda 4341$, visible.  For this
paper, we assume that any single emission line is one of the four
bright lines.

If a line is observed at $\lambda_o$, then we assume the galaxy's
redshift is one of $z_{\mathha} = \lambda_o / 6563~\rm{\AA} - 1$,
$z_{\mathoiii} = \lambda_o / 5007~\rm{\AA} - 1$, $z_{\mathhb} =
\lambda_o / 4861~\rm{\AA} - 1$, or $z_{\mathoii} = \lambda_o /
3727~\rm{\AA} - 1$.  We employ two independent methods, described
below, to assign probabilities to the four possible line identities or
redshifts: $P_{\mathha}$, $P_{\mathoiii}$, $P_{\mathhb}$, and
$P_{\mathoii}$.  Additionally, we set to zero the probabilities of
lines that satisfy the following conditions:

\begin{enumerate}
\item If the line is observed bluer than $0.98\lambda_{\mathha} =
  6431.5$~\AA, then $P_{\mathha} = 0$.  In other words, we do not
  permit blueshifts $z < -0.02$.
\item Of the $Q = 3$ and $Q = 4$ \deep\ galaxies at redshifts where
  both \ha\ and \hb\ are accessible, 80\% exhibit a Balmer decrement
  of $2.6 < \mathha/\mathhb < 7.7$ where the unit is spectral counts
  at the line peak.  Therefore, if the line is assumed to be \ha, and
  \hb\ also falls within the \deimos\ slit's spectral range, then
  $P_{\mathha} = 0$ if the smoothed spectral counts at the location of
  the single line are less than 2.6 or greater than 7.7 times the
  counts at the location of \hb\ within the errors of the photon
  counting statistics.
\item Similarly, 80\% of $Q = 3$ and $Q = 4$ \deep\ spectra with both
  [\oiii] lines exhibit an [\oiii] doublet ratio of $1.6 <
  [\mathoiii]~\lambda 5007/[\mathoiii]~\lambda 4959 < 5.0$.
  Therefore, if the line is assumed to be [\oiii] $\lambda 5007$ and
  [\oiii] $\lambda 4959$ should also be visible, then $P_{\mathoiii} =
  0$ if the counts at the location of [\oiii] $\lambda 5007$ are less
  than 1.6 or greater than 5.0 times the counts at the location of
  [\oiii] $\lambda 4959$ within the errors of the photon counting
  statistics.  The true physical ratio is always $[\mathoiii]~\lambda
  5007/[\mathoiii]~\lambda 4959 = 3$ \citep{ost06}.
\item In 80\% of $Q = 3$ and $Q = 4$ \deep\ spectra with both [\oiii]
  $\lambda 5007$ and \hb, $0.49 < [\mathoiii]/\mathhb < 4.2$.
  Therefore, $P_{\mathoiii} = 0$ if $[\mathoiii]/\mathhb > 4.2$ within
  the errors of the photon counting statistics.
\item Following condition (2), $P_{\mathhb} = 0$ if $\mathhb/\mathha >
  0.34$ within the errors of the photon counting statistics.
\item Following condition (4), $P_{\mathhb} = 0$ if
  $\mathhb/[\mathoiii] > 1.6$ within the errors of the photon counting
  statistics.
\item If the line drops to no visible counts within a window smaller
  than $220~\mathrm{km/s}$, the velocity separation of the [\oii]
  doublet, then $P_{\mathoii} = 0$ because the line cannot be
  dispersion-blended [\oii].
\end{enumerate}

\noindent Conditions (2) through (6) rely on an 80\% confidence
interval, which may seem strict, but the large majority of the
typically low signal-to-noise single lines pass these tests by virtue
of their large photon counting errors.

After all priors have been applied, we normalize the remaining
probabilities $P$ such that their sum is unity.

In the language of photo-$z$s, both of the following techniques are
``training set'' methods, or empirical calibrations.  Training methods
are immune to improper modeling and even incorrect photometric
zero-point offsets.  One weakness of training methods is that they
require large training sets for their accuracy and precision to
compare to that of modeling methods.

Another weakness is that they assume that the sample population
properties are similar to the training set properties.  For example,
if single-line galaxies preferentially have lower metallicity and
bluer colors than training set galaxies, then the methods described
here will search a skewed region of parameter space.  Furthermore, a
larger fraction of single-line galaxies show little or no continuum
than training set galaxies, which may skew colors.  For this paper, we
assume that the photometric properties of single-line galaxies are a
subset of the photometric properties of training set galaxies, meaning
that the parameter space around each single-line galaxy is
well-populated by training set galaxies at similar redshifts.

\subsection{Known-redshift sets}
\label{sec:known}
Both methods which assign probabilities to each of the four major
lines require sets of galaxies with well-measured redshifts.  The
known-redshift set consists of \deep\ targets with quality $Q = 3$ or
$4$ spectroscopic redshifts and with at least one emission line
detected at least $5\sigma$ above the noise.  There are 20,676 such
galaxies.  We identify the emission line containing the pixel with the
maximum counts in the spectrum smoothed in the same manner as the
single-line galaxy spectra in \S\,\ref{sec:sngls}.  For the remainder
of the training process, we treat the spectrum as containing only that
single line, but the redshift---and hence the identity of the single
line---is known.

Inaccurate redshifts in the known-redshift set of course lead to
inaccurate training and spurious line identification.  Repeat
observations have shown that fewer than 4.5\% of the 2,061 $Q = 3$
redshifts and fewer than 0.5\% of the 18,615 $Q = 4$ redshifts are
incorrect.  Therefore, the known-redshift set is less than 2\%
contaminated by incorrect redshifts.

Both of the following methods assume that the training set galaxies
populate the same observable parameter space as true single-line
galaxies.  The different target selection functions between Field 1
(6,532 galaxies) and Fields 2-4 (14,144 galaxies) make this point
especially important.  The underrepresentation of $z < 0.7$ galaxies,
which are present almost exclusively in Field 1, decreases the
certainty at which $z < 0.7$ galaxies may be identified.  However, as
long as $z < 0.7$ single-line galaxies populate a region of parameter
space distinct from $z > 0.7$ galaxies, and as long as $z < 0.7$
training set galaxies also populate that region, the following
algorithms should not mistake low-redshift galaxies for high-redshift
ones.  Therefore, we combine galaxies from all fields into one
training set.  We have also analyzed both methods with field
segregation.  We tested Field 1 galaxies using a training set with
only Field 1 galaxies, and we tested Fields 2-4 galaxies using a
training set with only Fields 2-4 galaxies.  The accuracy was
statistically indifferent from using a unified training set, but the
number of conclusively identified galaxies decreased slightly.

\subsection{Parameter space proximity method}
\label{sec:methoda}
We randomly divide the known-redshift set into training and evaluation
subsets.  The latter plays no role in the training, and we invoke it
only in \S\,\ref{sec:accuracy}.  A larger training set yields higher
precision whereas a larger evaluation set provides more confident
tests of the method.  We find that diverting 20\% of the
known-redshift set into the evaluation set, leaving 80\% for the
training set, gives a significant, untouched sample by which to judge
performance without significantly affecting precision.

The survey's three available broadband filter measurements, $B$, $R$,
and $I$, permit photo-$z$ measurements.  Because emission line
galaxies in different redshift ranges form different loci in $(B-R)$,
$(R-I)$, $R$, $\lambda_{\rm{observed}}$ space, a single galaxy's
position in that plane gives a guess at its redshift.  The two colors
alone provide enough information to determine some line identities,
but adding $R$ apparent magnitude increases the number of
identifications by $\sim 15\%$.  Observed wavelength also provides
discriminatory power because both the single line's observed
wavelength and the galaxy's broadband colors are functions of
redshift.

\begin{figure*}
\plottwo{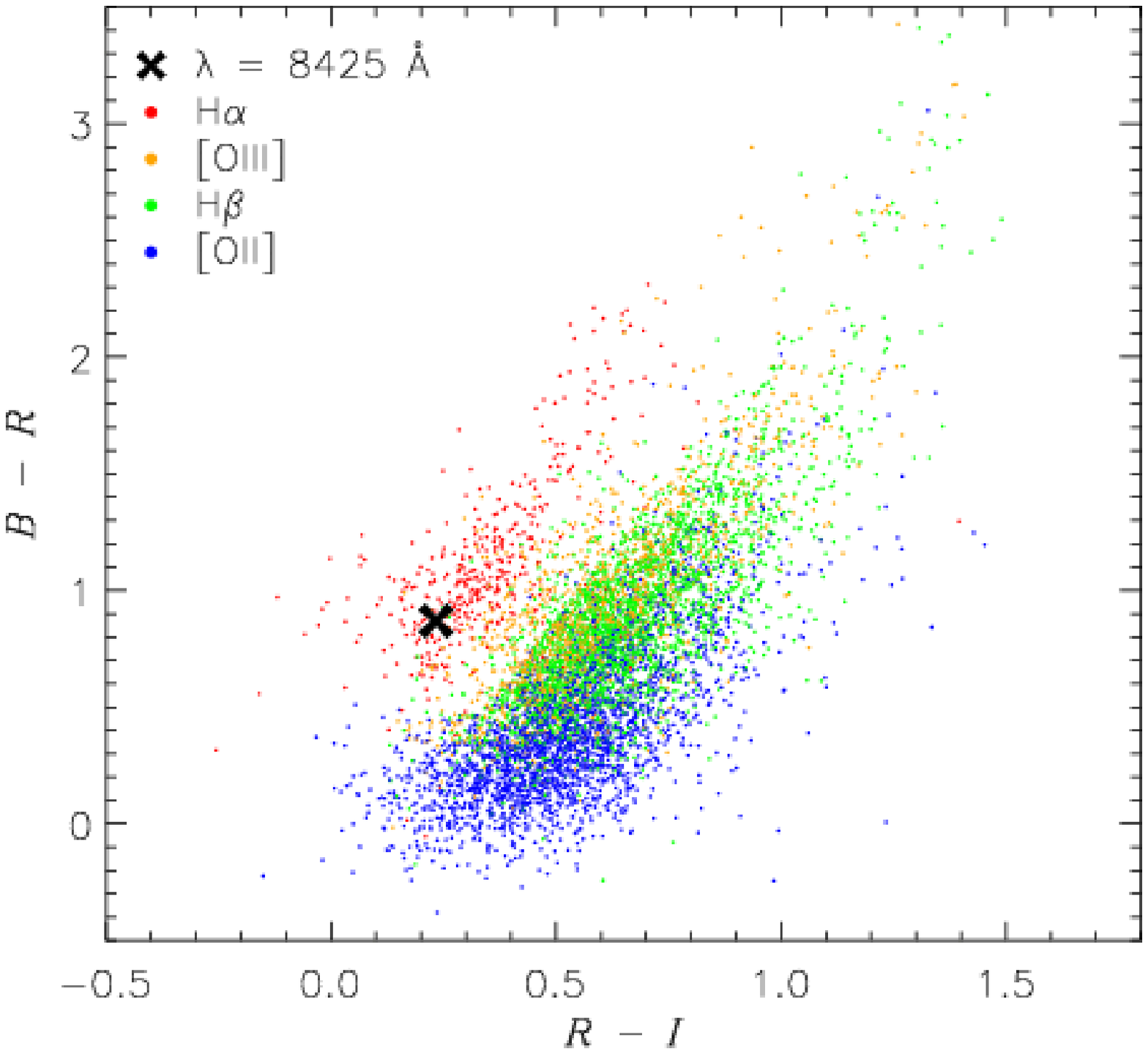}{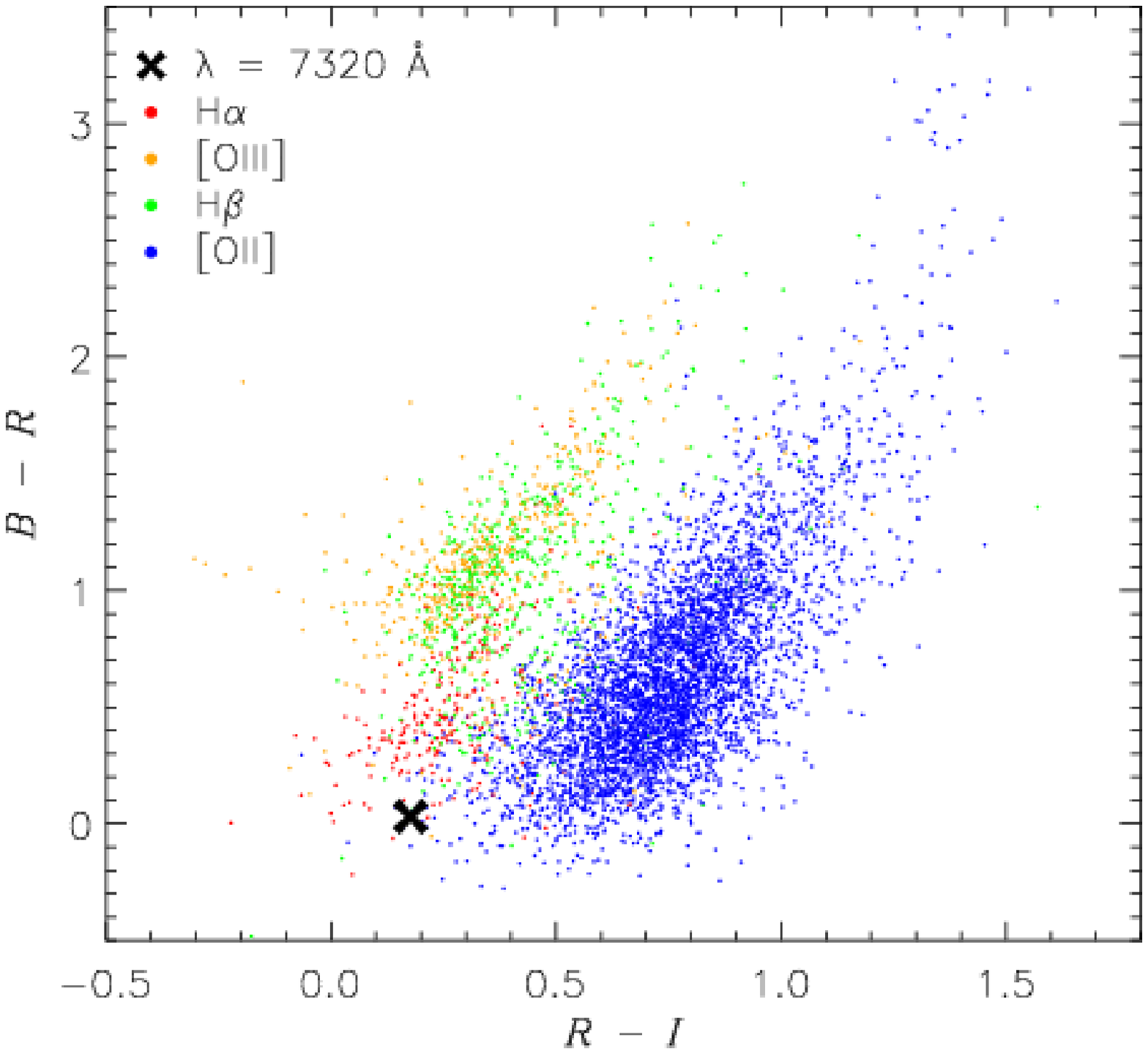}
\caption{\emph{Left:} A simplified representation of the PSP method,
  with the four-dimensional parameter space projected onto the $BRI$
  color-color plane.  Training set galaxies are color-coded by the
  values of their wavelength Gaussians (see \S\,\ref{sec:methoda}).
  Those galaxies whose \ha\ wavelength Gaussians exceed $5 \times
  10^{-4}$ are red, and so on for the other three line identities.  A
  galaxy with a single line observed at 8425~\AA\ ({\sf X}) falls in
  the middle of a locus of red points.  Therefore, this line is
  identified as \ha.  In reality, each training set point is a
  fixed-volume, four-dimensional Gaussian whose width is given by
  errors in observable quantities.  The sum of all the Gaussians in a
  particular line category at the location of the single-line galaxy
  is the probability that the single line also falls into that
  category.  The two suppressed axes are $R$ magnitude and observed
  wavelength. \emph{Right:} The same figure for a galaxy with
  different colors and wavelength.  In the absence of additional
  information, the identity of the single line cannot be determined
  because galaxies from multiple line identities populate the
  color-color space around the single-line galaxy.  \emph{Both
  panels:} The sharp color-color cut in Fields 2-4 ensures that only
  Field 1 galaxies populate the upper left corner of the diagram.
  Note that $z_{\mathha} = 0.28$ in the left panel and $z_{\mathha} =
  0.12$ in the right panel.  The second panel has fewer \ha\ points
  because the differential volume of the Universe is smaller at the
  second $z_{\mathha}$. \label{fig:methoda}}
\end{figure*}

We assign an identification confidence parameter $P_X^j$ for each
single-line galaxy, represented by index $j$.  $X$ represents one of
the bright lines.  Roughly, the confidence parameter is a measure of
distances from the point of the sample galaxy to each point in the
training set in parameter space.  More or closer points give a higher
$P_X^j$, and fewer or farther points give a lower $P_X^j$.  See
Fig.~\ref{fig:methoda} for a simplified representation of this method.

More precisely, every point in the training set is assigned a
four-dimensional Gaussian.  Each of the four axes corresponds to one
observable: $B-R$, $R-I$, $R$, and $\lambda_{\rm{observed}}$.  The
width of the Gaussian for each of the three color and magnitude
observables is given by the sum in quadrature of the photometric error
for the training set galaxy $i$ and the photometric error of the
single-line galaxy $j$.  The wavelength Gaussian is unique because it
distinguishes between the four line identities.  Each galaxy in the
training set has four wavelength Gaussians, each corresponding to one
of the four possible line identities, represented by $X$.  The
wavelength at the peak of Gaussian $X$ is $(1+z)\,\lambda_X$, where
$z$ is the known redshift of the training set galaxy and $\lambda_X$
is the rest wavelength of one of the four lines.  The width of this
Gaussian is the sum in quadrature of $\delta_\lambda = 1000 \, \delta
z \, \lambda_X$, where $\delta z$ is the error in the redshift from
the \texttt{spec1d} spectral template cross-correlation, and 106~\AA,
the median value of $\delta_\lambda$ for the evaluation set.  The
factor of 1000 is necessary to widen the Gaussians so that they
actually overlap.  It was chosen to optimize accuracy, but the results
are very insensitive to the precise value.  $P_X^j$ is the sum of the
values of all of these Gaussians corresponding to line $X$ at the
point where the single-line galaxy $j$ lies in the four-dimensional
parameter space.  In symbols,

\begin{eqnarray}
 P_{X}^{j} & = & \sum_{i} \frac{1}{ \sigma_{BR} \sigma_{RI}
 \sigma_{R} \sigma_{\lambda}} \exp \bigg\{ - \frac{\left[(B-R)_{i} -
 (B-R)_{j}\right]^{2}}{2 \sigma_{BR}^{2}} \label{eq:eta} \nonumber \\
 & & \quad - \frac{\left[(R-I)_{i} - (R-I)_{j}\right]^{2}}{2
 \sigma_{RI}^{2}} \nonumber \\ 
 & & \quad - \frac{\left[R_{i} -
 R_{j}\right]^{2}} {2 \sigma_{R}^{2}} - \frac{\left[(1+z_i)\lambda_{X}
 - \lambda_{j}\right]^{2}} {2 \sigma_{\lambda}^{2}} \bigg\} \\
 \sigma_{BR}^2 & \equiv & (\delta B_i)^{2} + (\delta R_i)^{2}
 + (\delta B_j)^{2} + (\delta R_j)^{2} \label{eq:sigmaBR} \\
 \sigma_{RI}^2 & \equiv & (\delta R_i)^{2} + (\delta I_i)^{2}
 + (\delta R_j)^{2} + (\delta I_j)^{2}\label{eq:sigmaRI} \\
 \sigma_{R}^2 & \equiv & (\delta R_i)^{2} + (\delta R_j)^{2}
 \label{eq:sigmaR} \\ 
 \sigma_{\lambda}^2 & \equiv & (1000 \, \delta z_i \, \lambda_{X})^2 +
 (106~\rm{\AA})^2 \label{eq:sigmalambda}
\end{eqnarray}

\noindent Although colors may be determined more precisely than
Eqs.~\ref{eq:sigmaBR} and \ref{eq:sigmaRI} suggest, we use the error
simply to broaden the Gaussians.  Additionally, the Gaussians of
galaxies with larger photometric errors will contribute less to
$P_X^j$ because they will have larger widths but the same volume.

The confidence parameters are subjected to the prior conditions
described above and normalized such that $\sum_X P_X^j = 1$.  This
normalization makes the Gaussian prefactor of $(2 \pi)^{-2}$
unnecessary.  Thus, $P_X^j$ represents the probability that the
identity of the single emission line in spectrum $j$ is $X$.

The criterion for a conclusive line identification is that $P_X^j$ for
a certain line $X$ exceeds $\rho_{\mathrm{PSP}}$, a tunable parameter.
A larger $\rho_{\mathrm{PSP}}$ will yield fewer conclusive line
identifications, but a larger fraction of them will be
correct.\footnote{When referring to line identification, ``accuracy''
in this paper means identifying lines correctly.  ``Precision'' means
the confidence with which a line is identified, or its value of $P_X$.
Therefore, increasing $\rho_{\mathrm{PSP}}$ imposes a stricter
condition on precision, thereby increasing accuracy.}

Because [\oiii] $\lambda 5007$ and \hb\ $\lambda 4861$ always fall so
close to each other in wavelength, it is very difficult to
discriminate between them.  However, it is possible to rule out \ha\
and [\oii] if $P_{\mathoiii}^j + P_{\mathhb}^j > \rho_{\mathrm{PSP}}$.
Additionally, conditions (3) through (6) described above may rule out
either [\oiii] $\lambda 5007$ or \hb, leaving only one possibility.

\subsection{Artificial neural network method}
An artificial neural network (ANN) can learn the functional
relationship between certain elements in a data set and derived
properties of the same set.  The first major implementation of ANNs in
astronomy was galaxy morphological classification \citep{sto92}, but
presently, the most common ANN implementation is photometric redshifts
\citep[e.g.,][]{fir03,van04}.  ANNs are more precise than other
machine learning methods \citep[e.g., trained decision tree
classifiers,][]{suc05} and model-independent, in contrast to template
fitting methods \citep[e.g.,][]{coe06,bro06}.  Given a set of
photometric data, such as broadband filter measurements, an ANN can
estimate a redshift and the error on that redshift.  This process
requires a very large training set for accurate and precise results.
Typically, a large known-redshift set with broadband data and
spectroscopic redshifts is divided randomly into three independent
sets: training, validation, and evaluation.  (In this paper, we divide
the known-redshift set into 60\% training, 20\% validation, and 20\%
evaluation to balance the precision a large training set affords with
the ability to test the ANN on an evaluation set.)  The ANN learns the
dependence of redshift on photometric observables from the training
set and interactively verifies its accuracy with the validation set.
After training and validation are complete, the ANN configuration is
fixed and unchangeable.  At this point, the evaluation set can test
how well the ANN has been trained (see \S\,\ref{sec:accuracy}).
Finally, the ANN may be applied to data without spectroscopic
redshifts to obtain photometric redshifts with an accuracy comparable
to that achieved with the evaluation set.

\begin{figure}
\plotone{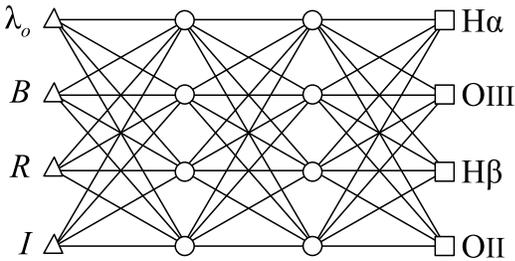}
\caption{The 4:4:4:4 artificial neural network architecture used for
  determining the identities of single emission lines.  Triangles
  represent the observed wavelength and three broadband filter inputs;
  squares represent the probabilities of each line identity; and
  circles represent hidden nodes. \label{fig:ann}}
\end{figure}

In determining single emission line identity, we employ an ANN with a
four-component output, using the publicly available code ANN$z$
\citep{col04}.  Each output corresponds to the probability of a line
identification.  During training, the brightest line in the spectrum
of a galaxy of known redshift is assigned a probability of 1 while the
other probabilities are 0.  In addition to providing $B$, $R$, and $I$
measurements as input to the ANN, we also provide the wavelength of
the single line.  Although there is danger in using irrelevant data as
input to an ANN, we justify the use of line wavelength by remarking
that the redshift and hence $BRI$ magnitudes and colors of a galaxy
would be different if the single line were \ha\ at observed wavelength
$\lambda_1$ rather than a different $\lambda_2$.  In full detail, the
ANN architecture is 4:4:4:4, meaning 4 inputs, 4 outputs, and 2 hidden
layers with 4 nodes each, shown diagrammatically in
Fig.~\ref{fig:ann}.  We find that other similar architectures do not
affect line identification significantly.

In determining the functional relationship between inputs and outputs,
an ANN finds the optimal configuration of weights to connect each
node.  Before training begins, the ANN is initialized with random
weights.  For this reason, ANNs with the same architecture and same
training sets but with different initializations will usually find
different local minima.  The solution is to use the average of the
results of a committee of separately initialized ANNs.  We use
committees of 20 ANNs identical in every way except for their
initially randomized weights.

In analogy to the PSP method, we call the ANN outputs $P_X^j$, which
are subjected to the seven conditions described in
\S\,\ref{sec:methods}.  Line $j$ is identified conclusively only if a
single $P_X^j$ or $P_{\mathoiii}^j + P_{\mathhb}^j$ exceeds
$\rho_{\mathrm{ANN}}$.

\section{Accuracy}
\label{sec:accuracy}

Although setting aside evaluation sets reduces the training set sizes
and hence reduces precision, evaluation establishes confidence in the
line identifications.  In \S\,\ref{sec:known}, we identified one line
in each known-redshift galaxy to serve as the surrogate ``single''
line.  In this section, we subject each evaluation galaxy to both
methods and report the estimated identity of each line.

\begin{deluxetable}{lrrrr|r}
\tablecolumns{6}
\tablewidth{0pc}
\tablecaption{Evaluation set accuracy, all fields.
\label{tab:acc}}
\tablehead{\multicolumn{6}{c}{PSP Method} \\
\hline
\colhead{} & \multicolumn{4}{c}{Actual Line Identity} & \colhead{} \\
\cline{2-5}
\colhead{Identified As} & \colhead{\ha} & \colhead{[\oiii]} & 
\colhead{\hb} & \colhead{[\oii]} & \colhead{Total}}
\startdata
\ha  &  \textbf{201}  &  15  &  2  &  2  &  220  \\
{[\oiii]}  &  14  &  \textbf{200}  &  \textbf{8}  &  4  &  226  \\
\hb  &  3  &  \textbf{36}  &  \textbf{40}  &  4  &  83  \\
{[\oii]}  &  0  &  0  &  0  &  \textbf{1896}  &  1896  \\
{[\oiii]} or \hb  &  28  &  \textbf{571}  &  \textbf{365}  &  25  &  989  \\
Inconclusive  &  63  &  79  &  9  &  567  &  718  \\
\hline
Total  &  309  &  901  &  424  &  2498  &  4132  \\
\hline\hline \\
\multicolumn{6}{c}{ANN Method} \\
\hline
\colhead{} & \multicolumn{4}{c}{Actual Line Identity} & \colhead{} \\
\cline{2-5}
\colhead{Identified As} & \colhead{\ha} & \colhead{[\oiii]} & 
\colhead{\hb} & \colhead{[\oii]} & \colhead{Total} \\
\hline
\ha  &  \textbf{226}  &  11  &  2  &  1  &  240  \\
{[\oiii]}  &  14  &  \textbf{545}  &  \textbf{32}  &  5  &  596  \\
\hb  &  0  &  \textbf{18}  &  \textbf{60}  &  1  &  79  \\
{[\oii]}  &  0  &  0  &  0  &  \textbf{2184}  &  2184  \\
{[\oiii]} or \hb  &  8  &  \textbf{217}  &  \textbf{316}  &  16  &  557  \\
Inconclusive  &  61  &  110  &  14  &  291  &  476  \\
\hline
Total  &  309  &  901  &  424  &  2498  &  4132  \\
\enddata
\end{deluxetable}

\begin{deluxetable}{lrrrr|r}
\tablecolumns{6}
\tablewidth{0pc}
\tablecaption{Evaluation set accuracy, Field 1.
\label{tab:acc1}}
\tablehead{\multicolumn{6}{c}{PSP Method} \\
\hline
\colhead{} & \multicolumn{4}{c}{Actual Line Identity} & \colhead{} \\
\cline{2-5}
\colhead{Identified As} & \colhead{\ha} & \colhead{[\oiii]} & 
\colhead{\hb} & \colhead{[\oii]} & \colhead{Total}}
\startdata
\ha  &  \textbf{166}  &  8  &  2  &  2  &  178  \\
{[\oiii]}  &  12  &  \textbf{106}  &  \textbf{6}  &  1  &  125  \\
\hb  &  0  &  \textbf{20}  &  \textbf{17}  &  0  &  37  \\
{[\oii]}  &  0  &  0  &  0  &  \textbf{372}  &  372  \\
{[\oiii]} or \hb  &  19  &  \textbf{223}  &  \textbf{133}  &  7  &  382  \\
Inconclusive  &  50  &  62  &  4  &  96  &  212  \\
\hline
Total  &  247  &  419  &  162  &  478  &  1306  \\
\hline\hline \\
\multicolumn{6}{c}{ANN Method} \\
\hline
\colhead{} & \multicolumn{4}{c}{Actual Line Identity} & \colhead{} \\
\cline{2-5}
\colhead{Identified As} & \colhead{\ha} & \colhead{[\oiii]} & 
\colhead{\hb} & \colhead{[\oii]} & \colhead{Total} \\
\hline
\ha  &  \textbf{179}  &  5  &  0  &  1  &  185  \\
{[\oiii]}  &  8  &  \textbf{215}  &  \textbf{15}  &  0  &  238  \\
\hb  &  0  &  \textbf{11}  &  \textbf{22}  &  1  &  34  \\
{[\oii]}  &  0  &  0  &  0  &  \textbf{432}  &  432  \\
{[\oiii]} or \hb  &  5  &  \textbf{105}  &  \textbf{113}  &  3  &  226  \\
Inconclusive  &  55  &  83  &  12  &  41  &  191  \\
\hline
Total  &  247  &  419  &  162  &  478  &  1306  \\
\enddata
\end{deluxetable}

\begin{deluxetable}{lrrrr|r}
\tablecolumns{6}
\tablewidth{0pc}
\tablecaption{Evaluation set accuracy, Fields 2-4.
\label{tab:acc234}}
\tablehead{\multicolumn{6}{c}{PSP Method} \\
\hline
\colhead{} & \multicolumn{4}{c}{Actual Line Identity} & \colhead{} \\
\cline{2-5}
\colhead{Identified As} & \colhead{\ha} & \colhead{[\oiii]} & 
\colhead{\hb} & \colhead{[\oii]} & \colhead{Total}}
\startdata
\ha  &  \textbf{35}  &  7  &  0  &  0  &  42  \\
{[\oiii]}  &  2  &  \textbf{94}  &  \textbf{2}  &  3  &  101  \\
\hb  &  3  &  \textbf{16}  &  \textbf{23}  &  4  &  46  \\
{[\oii]}  &  0  &  0  &  0  &  \textbf{1524}  &  1524  \\
{[\oiii]} or \hb  &  9  &  \textbf{348}  &  \textbf{232}  &  18  &  607  \\
Inconclusive  &  13  &  17  &  5  &  471  &  506  \\
\hline
Total  &  62  &  482  &  262  &  2020  &  2826  \\
\hline\hline \\
\multicolumn{6}{c}{ANN Method} \\
\hline
\colhead{} & \multicolumn{4}{c}{Actual Line Identity} & \colhead{} \\
\cline{2-5}
\colhead{Identified As} & \colhead{\ha} & \colhead{[\oiii]} & 
\colhead{\hb} & \colhead{[\oii]} & \colhead{Total} \\
\hline
\ha  &  \textbf{40}  &  5  &  2  &  0  &  47  \\
{[\oiii]}  &  6  &  \textbf{334}  &  \textbf{16}  &  3  &  359  \\
\hb  &  0  &  \textbf{11}  &  \textbf{41}  &  0  &  52  \\
{[\oii]}  &  0  &  0  &  0  &  \textbf{1751}  &  1751  \\
{[\oiii]} or \hb  &  5  &  \textbf{105}  &  \textbf{202}  &  19  &  331  \\
Inconclusive  &  11  &  27  &  1  &  247  &  286  \\
\hline
Total  &  62  &  482  &  262  &  2020  &  2826  \\
\enddata
\end{deluxetable}

We choose $\rho_{\mathrm{PSP}} = 0.9$ and $\rho_{\mathrm{ANN}} = 0.8$,
motivated by the arguments below.  Tables
\ref{tab:acc}-\ref{tab:acc234} detail the results separated by field.
Each column is an actual surrogate single line identity, and each row
is a result from the identification algorithm.  Correct
identifications are shown in bold, where we consider the [\oiii], \hb,
and ``[\oiii] or \hb'' categories to be correct for both [\oiii] and
\hb\ lines.  (If we assume the rest wavelength of the single line is
the average of 4861~\AA\ and 5007~\AA, then the redshift will be
skewed by 1.5\%, which is more precise than even the best photo-$z$s.)

The prior conditions in \S\,\ref{sec:methods} improve the results
significantly.  The priors corrected 465 identifications in the PSP
method, mostly [\oiii] and [\oii], and 597 identifications in the ANN
method, overwhelming [\oiii].  The large majority of the
identifications corrected to \ha\ and [\oii] were inconclusive before
the application of the prior conditions.  Those corrected to [\oiii]
were mostly ``[\oiii] or \hb'' or inconclusive, and those corrected to
\hb\ were mostly ``[\oiii] or \hb.''  The prior conditions work for
many [\oiii] lines in the ANN method because the $P_{\mathoiii}$
values for those lines are high even before the application of the
prior conditions.  In the PSP method, even though the prior conditions
can eliminate \hb\ as a possibility, $P_{\mathoiii}$ is not large
enough for a conclusive identification.  Our choice of a smaller
$\rho_{\mathrm{ANN}}$ than $\rho_{\mathrm{PSP}}$ creates this
particular success of the ANN method.

Both methods show no large statistical difference between the fields
with different selection functions.  Field 1 contains more actual \ha\
lines than Fields 2-4, but the accuracy with which they are identified
is about the same.  This similarity between the fields justifies using
one known-redshift set instead of one for Field 1 and a different one
for Fields 2-4.

The ANN method is superior to the PSP method in both accuracy and
number of conclusive identifications.  The PSP method in particular
identifies 9.1\% of the actual \ha\ lines as ``[\oiii] or \hb''
compared to only 2.6\% for the ANN method.  Interestingly, 8.6\% and
5.8\% of the lines identified as \ha\ in the PSP and ANN methods are
incorrect, despite the underrepresentation of \ha\ lines in Fields
2-4.  Not surprisingly, the PSP method performs poorly at identifying
[\oiii] and \hb.  However, the ANN method performs very well at
identifying [\oiii] (91.4\% correct identifications) because the prior
conditions correct so many identifications to [\oiii].  Every line
classified as [\oii] in both methods is correct, largely because of
condition (7) in \S\,\ref{sec:methods}.  The ``[\oiii] or \hb''
category is 94.6\% and 95.7\% accurate for the PSP and ANN methods
respectively, but the ANN method conclusively and correctly identifies
many more lines as [\oiii].

\begin{deluxetable}{lrrrrrr|r}
\tablecolumns{8}
\tablewidth{0pc}
\tablecaption{Comparison between PSP and ANN methods on evaluation
  set. \label{tab:eval_compare}}
\tablehead{
\colhead{} & \multicolumn{6}{c}{ANN Method} & \colhead{} \\
\cline{2-7}
\colhead{PSP Method} & \colhead{\ha} & \colhead{[\oiii]} & 
\colhead{\hb} & \colhead{[\oii]} & \colhead{[\oiii]/\hb} &
\colhead{Inc.} & \colhead{Total}}
\startdata
\ha  &  \textbf{182}  &  3  &  0  &  0  &  3  &  32  &  220  \\
{[\oiii]}  &  5  &  \textbf{149}  &  0  &  0  &  42  &  30  &  226  \\
\hb  &  4  &  12  &  \textbf{26}  &  1  &  28  &  12  &  83  \\
{[\oii]}  &  0  &  0  &  0  &  \textbf{1839}  &  1  &  56  &  1896  \\
{[\oiii]} or \hb  &  11  &  411  &  52  &  1  &  \textbf{464}  &  50  &  989  \\
Inconclusive  &  38  &  21  &  1  &  343  &  19  &  \textbf{296}  &  718  \\
\hline
Total  &  240  &  596  &  79  &  2184  &  557  &  476  &  4132  \\
\enddata
\end{deluxetable}

Table \ref{tab:eval_compare} shows the number of lines from the
evaluation set that were identified the same and differently between
the two methods.  The numbers of lines that fell into the same
category in both methods are shown in bold.  Very few conclusive
identifications are different between the two methods.


\begin{figure*}
 \plottwo{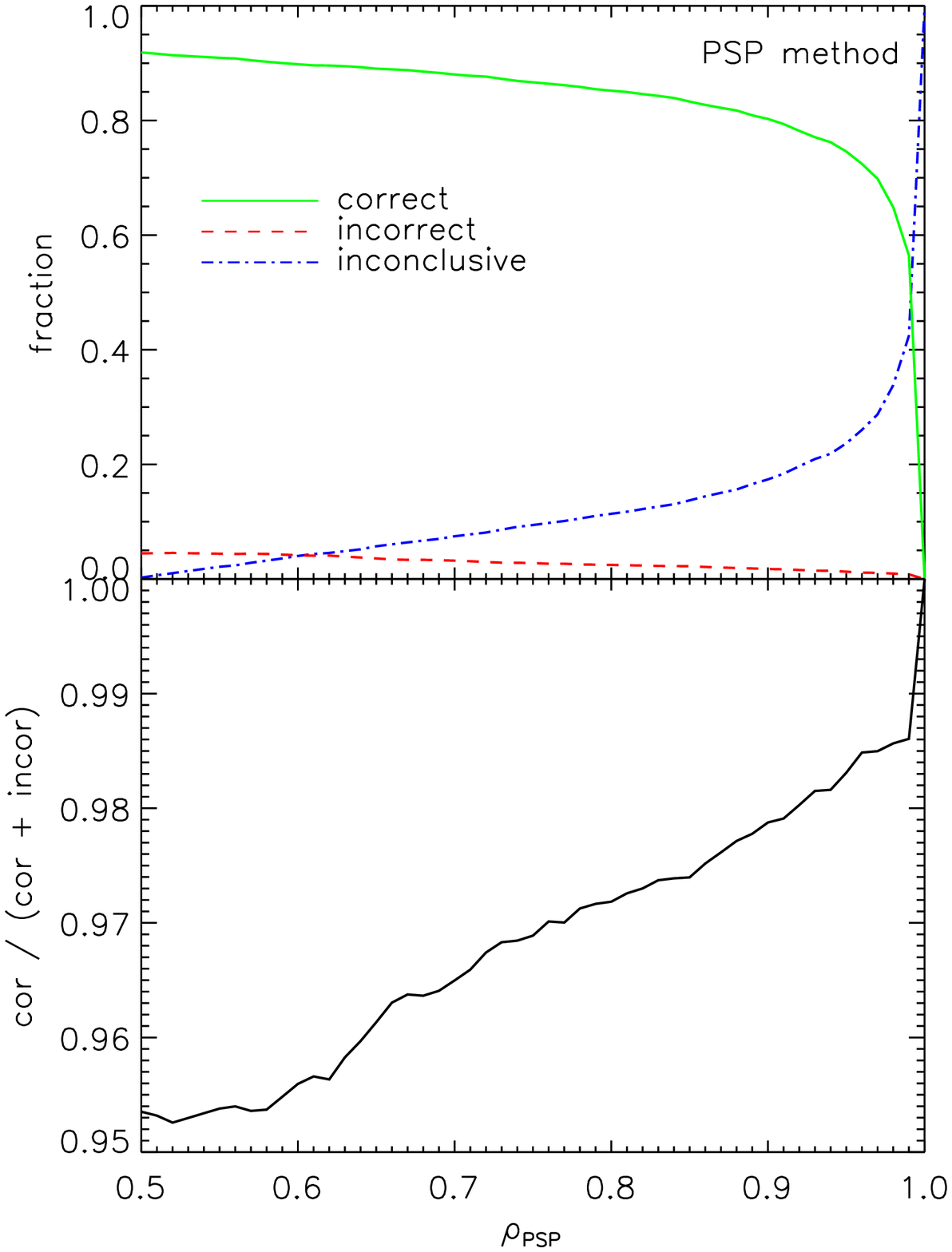}{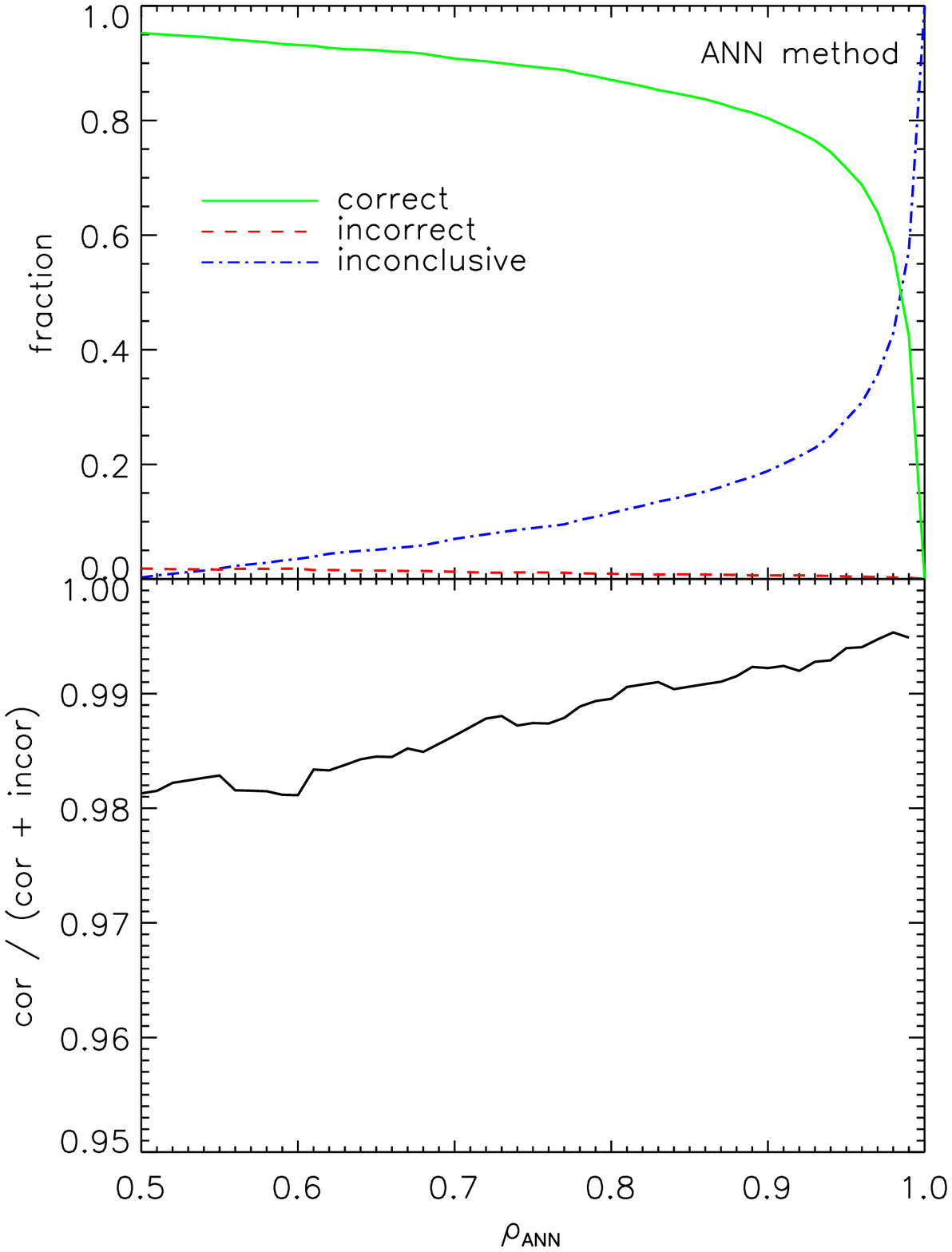}
 \caption{\emph{Top:} Fraction of total evaluation set galaxies
 correctly, incorrectly, and inconclusively identified in both methods
 versus the accuracy parameter $\rho$.  We allow the [\oiii], \hb, and
 ``[\oiii] or \hb'' identifications to be ``correct'' as long as the
 actual line is [\oiii] or \hb.  \emph{Bottom:} The ratio of correct
 identifications to the sum of correct and incorrect identifications,
 as defined in the top panel. \label{fig:acc}}
\end{figure*}


We find the optimal accuracy parameter $\rho$ for each method by
attempting to maximize accuracy while minimizing the number of
inconclusive identifications.  We classify accurate \ha, [\oii], and
``[\oiii] or \hb'' identifications as ``correct.''  Individual [\oiii]
and \hb\ identifications are correct if the actual line is either
[\oiii] or \hb.  Lines where $P_X^j < \rho$ for all $X$ are
``inconclusive'' except where $P_{\mathoiii}^j + P_{\mathhb}^j >
\rho$.  Inaccurate \ha, [\oii], and ``[\oiii] or \hb'' identifications
are ``incorrect,'' as well as individual [\oiii] and \hb\
identifications where the actual line is \ha\ or [\oii].  Figure
\ref{fig:acc} shows the results for each method.  We find that
$\rho_{\mathrm{PSP}} = 0.9$ and $\rho_{\mathrm{ANN}} = 0.8$ give
ratios of correct to the sum of correct and incorrect results near
0.98 without sacrificing many conclusive identifications.

\section{Results on Single-Line Galaxies}
\label{sec:results}
\subsection{Single Line Identification}

\begin{deluxetable}{lrr|r}
\tablecolumns{4}
\tablewidth{0pc}
\tablecaption{Single line identification.
\label{tab:sngl}}
\tablehead{\multicolumn{4}{c}{PSP Method} \\
\hline
\colhead{Identified As} & \colhead{Field 1} & \colhead{Fields 2-4} & \colhead{All Fields}}
\startdata
\ha  &  51  &  60  &  111  \\
{[\oiii]}  &  16  &  16  &  32  \\
\hb  &  12  &  13  &  25  \\
{[\oii]}  &  9  &  50  &  59  \\
{[\oiii]} or \hb  &  85  &  89  &  174  \\
Inconclusive  &  91  &  148  &  239  \\
\hline
Total  &   264  &   376  &   640  \\
\hline\hline \\
\multicolumn{4}{c}{ANN Method} \\
\hline
\colhead{Identified As} & \colhead{Field 1} & \colhead{Fields 2-4} & \colhead{All Fields} \\
\hline
\ha  &  79  &  87  &  166  \\
{[\oiii]}  &  55  &  47  &  102  \\
\hb  &  4  &  11  &  15  \\
{[\oii]}  &  21  &  73  &  94  \\
{[\oiii]} or \hb  &  40  &  55  &  95  \\
Inconclusive  &  65  &  103  &  168  \\
\hline
Total  &   264  &   376  &   640  \\
\enddata
\end{deluxetable}

We apply both methods to the 640 truly single emission lines.
Table~\ref{tab:sngl} lists the fraction of single lines identified in
each category separated by field.  A comparison of each column in this
table to the ``Total'' columns in
Tables~\ref{tab:acc}-\ref{tab:acc234} immediately shows that the
single-line population contains fewer [\oii] lines because [\oii] is
often resolved and identifiable in \deep\ spectra.  It can be a single
line only when the galaxy's internal velocities blend the doublet.

\begin{figure*}
 \plottwo{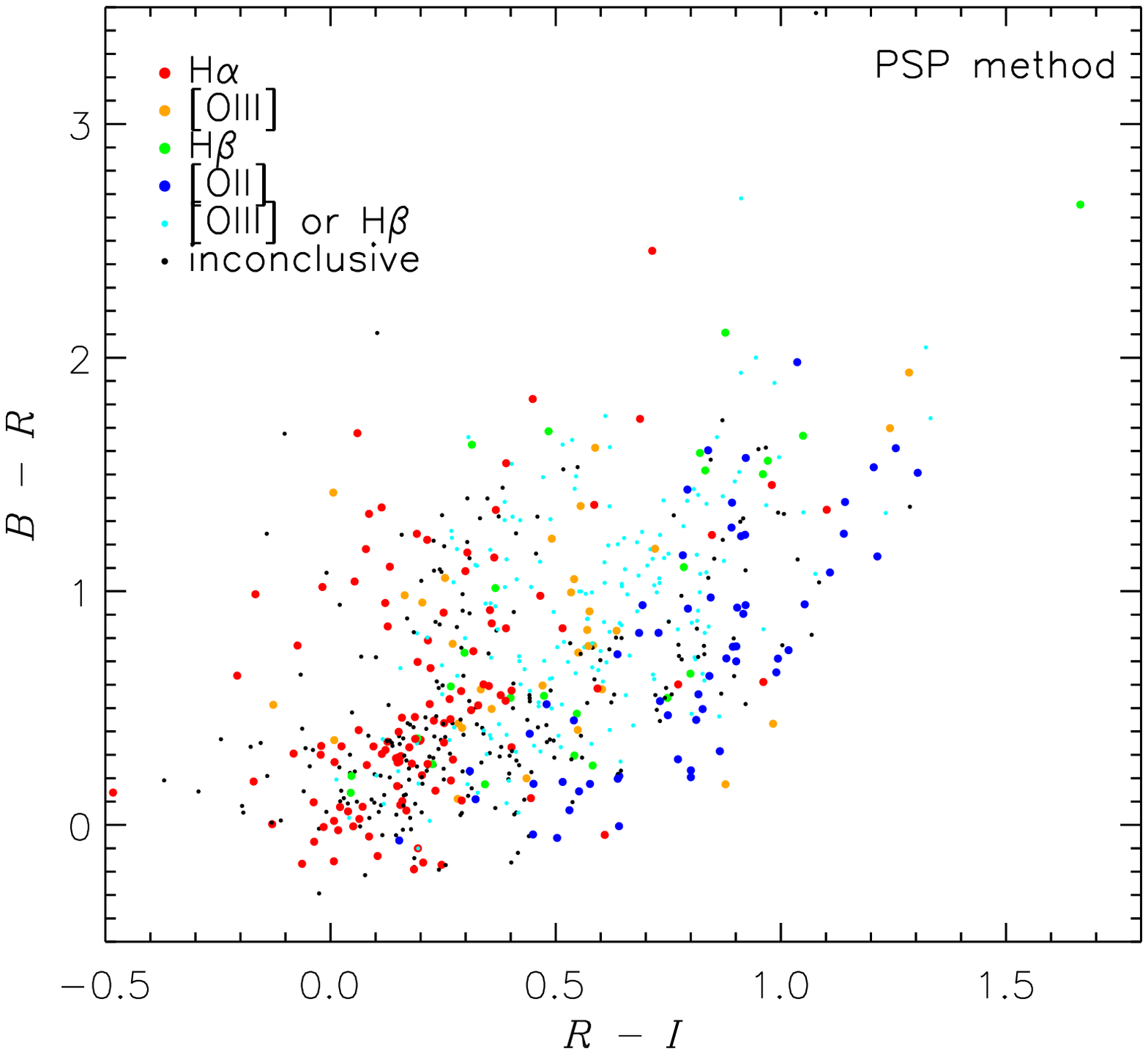}{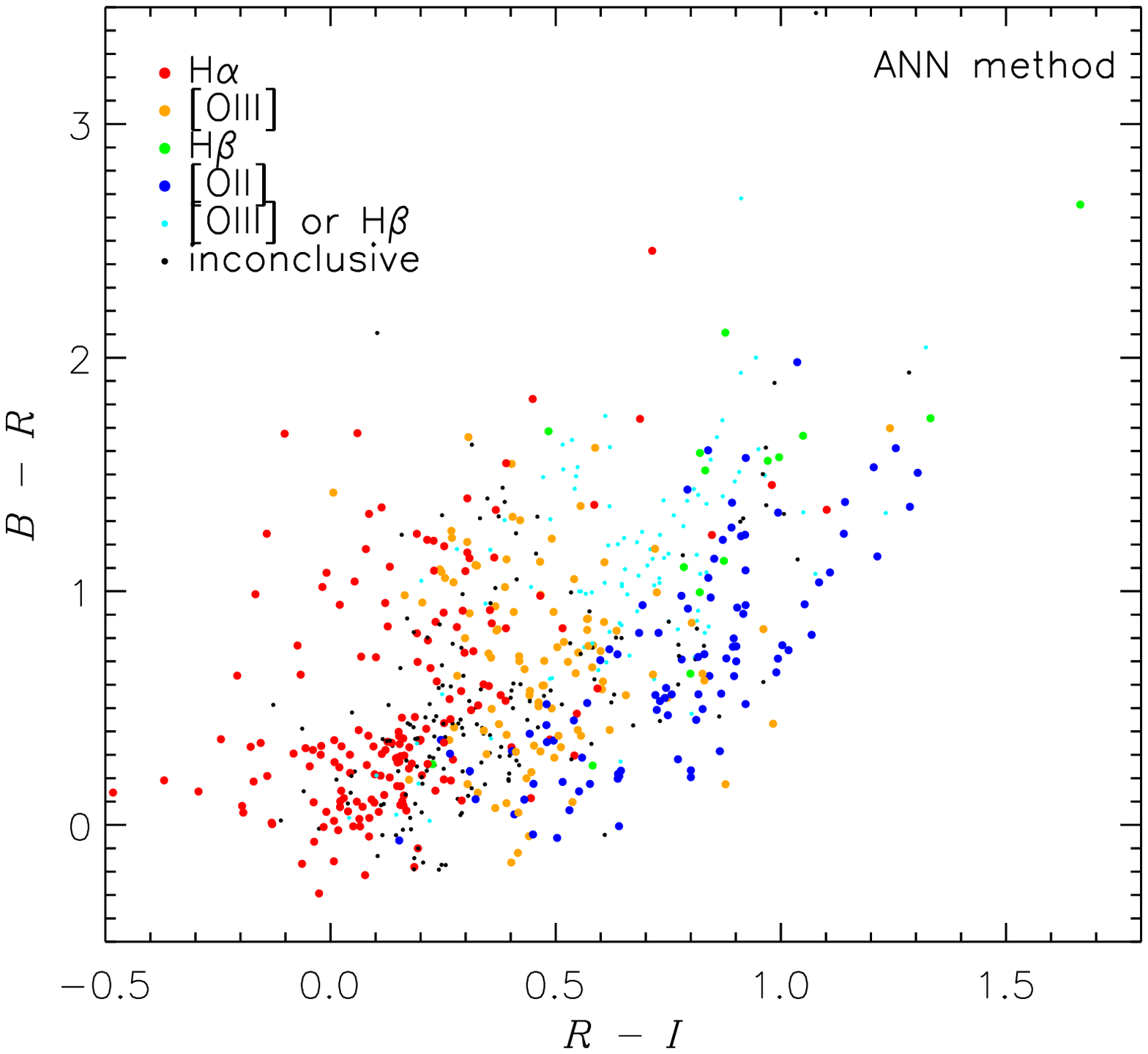}
 \caption{$BRI$ color-color diagrams for the 640 single-line galaxies
 in all four fields color-coded by their identifications in both
 methods.  Points that represent the ``[\oiii] or \hb'' and
 ``inconclusive'' identifications are smaller so that they do not
 obscure the other points. \label{fig:bri}}
\end{figure*}

Of the four observables used to determine line identities, the two
colors are the most powerful.  Figure \ref{fig:bri} shows the $BRI$
color-color plane for the lines identified in both methods.  It is
useful to compare this figure with Fig.~\ref{fig:methoda}.  Conclusive
line identifications are possible in regions where the training set
categories overlap because $R$ magnitude and observed wavelength also
help to determine line identity and, more importantly, because the
seven conditions described in \S\,\ref{sec:methods} rule out certain
line identities, leaving one dominant identity.

In contrast to the evaluation set, the priors change a large fraction
of the single-line set identifications.  In the PSP method, the priors
changed the identifications of 517 (80.8\%) of the single lines,
largely from \hb\ to ``[\oiii] or \hb'' or from [\oii] to
inconclusive.  In the ANN method, priors changed the identifications
of 285 (44.5\%) of the single lines, mostly from [\oii] or
inconclusive to \ha, [\oiii], or \hb.  While these large fractions may
seem to diminish the power of the core PSP and ANN methods, the priors
eliminate only one line in most cases.  It is still up to the core
algorithm to choose among the remaining three lines based on broadband
photometry and observed wavelength.

\begin{deluxetable}{lrrrrrr|r}
\tablecolumns{8}
\tablewidth{0pc}
\tablecaption{Comparison between PSP and ANN methods on single-line
  galaxies. \label{tab:sngl_compare}}
\tablehead{
\colhead{} & \multicolumn{6}{c}{ANN Method} & \colhead{} \\
\cline{2-7}
\colhead{PSP Method} & \colhead{\ha} & \colhead{[\oiii]} & 
\colhead{\hb} & \colhead{[\oii]} & \colhead{[\oiii]/\hb} &
\colhead{Inc.} & \colhead{Total}}
\startdata
\ha  &  \textbf{100}  &  0  &  0  &  0  &  0  &  11  &  111  \\
{[\oiii]}  &  1  &  \textbf{22}  &  0  &  0  &  0  &  9  &  32  \\
\hb  &  3  &  1  &  \textbf{11}  &  0  &  0  &  10  &  25  \\
{[\oii]}  &  0  &  0  &  0  &  \textbf{58}  &  0  &  1  &  59  \\
{[\oiii]} or \hb  &  1  &  61  &  4  &  0  &  \textbf{87}  &  21  &  174  \\
Inconclusive  &  61  &  18  &  0  &  36  &  8  &  \textbf{116}  &  239  \\
\hline
Total  &  166  &  102  &  15  &  94  &  95  &  168  &  640  \\
\enddata
\end{deluxetable}

Table \ref{tab:sngl_compare} shows the number of single lines that
were identified the same and differently between the two methods.
Conclusive line identifications do not change significantly between
the two methods.

\begin{deluxetable}{lrrrr|r}
\tablecolumns{6}
\tablewidth{0pc}
\tablecaption{Recovered redshifts.
\label{tab:recovered}}
\tablehead{\multicolumn{6}{c}{PSP Method} \\
\hline
\colhead{} & \multicolumn{4}{c}{Actual Line Identity} & \colhead{} \\
\cline{2-5}
\colhead{Identified As} & \colhead{\ha} & \colhead{[\oiii]} & 
\colhead{\hb} & \colhead{[\oii]} & \colhead{Total}}
\startdata
\ha  &  \textbf{5}  &  0  &  0  &  0  &  5  \\
{[\oiii]}  &  0  &  \textbf{2}  &  \textbf{0}  &  0  &  2  \\
\hb  &  0  &  \textbf{0}  &  \textbf{0}  &  0  &  0  \\
{[\oii]}  &  0  &  0  &  0  &  \textbf{9}  &  9  \\
{[\oiii]} or \hb  &  1  &  \textbf{3}  &  \textbf{3}  &  1  &  8  \\
Inconclusive  &  2  &  5  &  0  &  2  &  9  \\
\hline
Total  &  8  &  10  &  3  &  12  &  33  \\
\hline\hline \\
\multicolumn{6}{c}{ANN Method} \\
\hline
\colhead{} & \multicolumn{4}{c}{Actual Line Identity} & \colhead{} \\
\cline{2-5}
\colhead{Identified As} & \colhead{\ha} & \colhead{[\oiii]} & 
\colhead{\hb} & \colhead{[\oii]} & \colhead{Total} \\
\hline
\ha  &  \textbf{6}  &  1  &  0  &  0  &  7  \\
{[\oiii]}  &  0  &  \textbf{4}  &  \textbf{0}  &  0  &  4  \\
\hb  &  0  &  \textbf{0}  &  \textbf{0}  &  0  &  0  \\
{[\oii]}  &  0  &  0  &  0  &  \textbf{10}  &  10  \\
{[\oiii]} or \hb  &  0  &  \textbf{2}  &  \textbf{3}  &  0  &  5  \\
Inconclusive  &  2  &  3  &  0  &  2  &  7  \\
\hline
Total  &  8  &  10  &  3  &  12  &  33  \\
\enddata
\end{deluxetable}

During the visual inspection of all single-line candidates, we
identified the redshifts of 33 galaxies using additional lines that
previous inspectors missed because they were dim or nearly hidden.
Table \ref{tab:recovered} summarizes the results.  The PSP method
identifies 22 lines correctly, 2 incorrectly, and 9 inconclusively.
The ANN method identifies 25 lines correctly, 1 incorrectly, and 7
inconclusively.  Because these galaxies with recovered redshifts are
so similar to the other single-line galaxies, their high success rate
lends credence to both algorithms.

\subsection{Spectral Coaddition}
\label{sec:coadd}
A useful statistical check is to coadd the one-dimensional spectra of
all of the galaxies identified in each category.  We shift all the
spectra to their rest frames and normalize them such that the median
number of counts in a pixel is 1.  Then, we coadd them with inverse
variance weighting.  The coadded spectra are smoothed with a Gaussian
window function of $\sigma = 2.5$~pixels to approximate the
instrumental resolution.  If a line is identified properly, other
spectral features associated with the single line should emerge above
the noise.  On the other hand, if the line is misidentified, some
spectral features associated with other lines may be present.
Although this technique cannot identify individual lines, it can give
an idea of overall success or failure for each identification
category.


\begin{figure*}
 \centering
 \includegraphics[angle=270,width=0.85\textwidth]{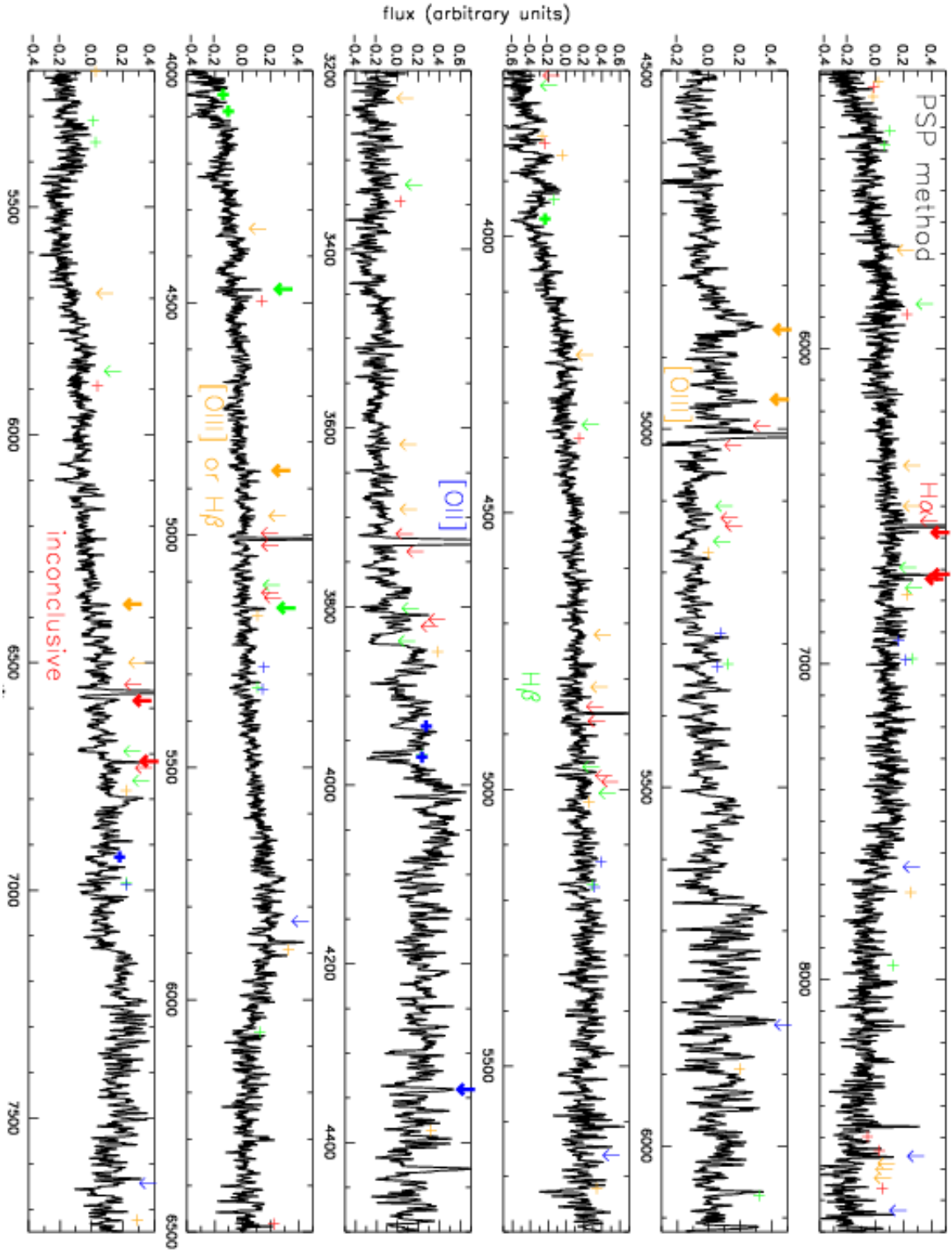}
 \caption{The coadded spectra of lines identified in each of the six
 categories.  Each spectrum is labeled in Angstroms and in the rest
 frame of the identified line.  ``[\oiii] or \hb'' is in the rest
 frame of [\oiii] $\lambda 5007$, and ``inconclusive'' is in the rest
 frame of \ha\ $\lambda 6563$.  The symbols indicate spectral features
 at four different redshifts: red if the single line is \ha, orange
 for [\oiii], green for \hb, and blue for [\oii].  Arrows represent
 emission lines, and pluses represent absorption lines.  Bold symbols
 draw attention to noticeable features, and thin symbols mark the
 locations of absent features.  The peaks of the single lines are
 stronger than the maximum plotted flux.
 \label{fig:coadd}}
\end{figure*}


\setcounter{figure}{4}

\begin{figure*}
 \centering
 \includegraphics[angle=270,width=0.85\textwidth]{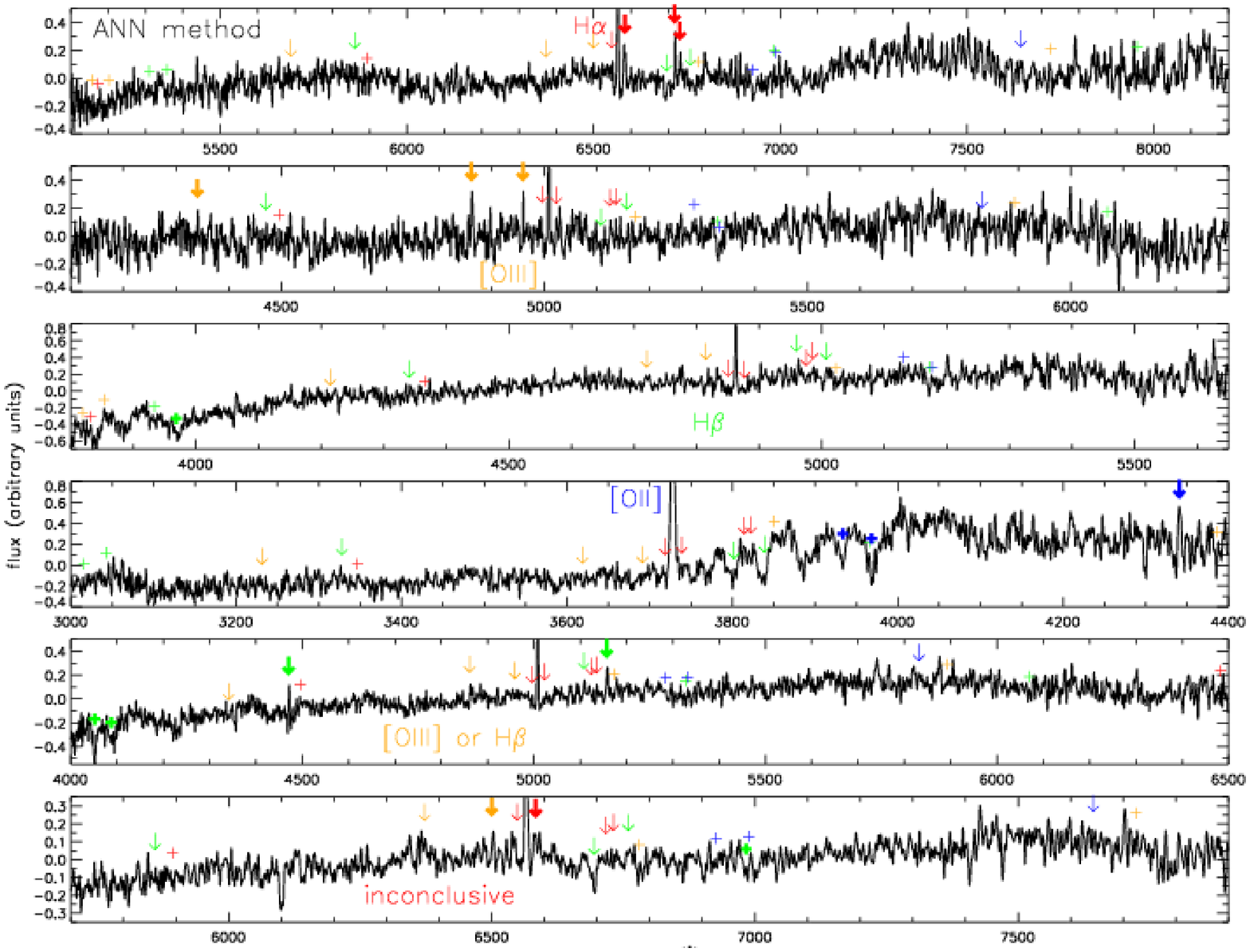}
\caption{continued}
\end{figure*}


Figure~\ref{fig:coadd} shows the coadded spectra for the galaxies
identified in each of the six categories.  Emission and absorption
features shown at four different redshifts---corresponding to the four
allowed identities of each single line---are marked on each spectrum.
The fourteen features are [\oii] $\lambda 3727$, CaH $\lambda 3934$,
CaK $\lambda 3969$, \hg\ $\lambda 4341$, \hb\ $\lambda 4861$, [\oiii]
$\lambda\lambda 4959$, 5007, Mgb $\lambda 5173$, NaD $\lambda 5893$,
[\nii] $\lambda\lambda 6548$, 6583, \ha\ $\lambda 6563$, and [\sii]
$\lambda\lambda 6716$, 6731.  Bold symbols direct attention to
potentially real spectral features.

Both methods show [\nii] and [\sii] in the \ha\ spectra; [\oiii]
$\lambda 4959$ and \hb\ in the [\oiii] spectra; CaK in the \hb\
spectra; and CaH, CaK, and \hg\ as well as high-order Balmer
absorption in the [\oii] spectra.  The [\oii] lines are broad, as
expected.  The ``[\oiii] or \hb'' spectra for both methods display
CaH, CaK, \hg, and [\oiii] $\lambda 5007$ associated with \hb.  The
PSP method spectrum may also contain [\oiii] $\lambda 4959$ associated
with [\oiii] $\lambda 5007$.  Finally, the inconclusive category
contains few convincing lines, but features associated with \ha\ and
[\oiii] may be present.  The presence of expected lines and absence of
others in conclusively identified single-line galaxy spectra
strengthens credibility in the identification of single emission
lines.

\section{Conclusions}
\label{sec:conclusions}
The methods presented here combine photometry, spectroscopy, and
physically motivated arguments about line flux ratios to determine the
identity of single emission lines in galaxy spectra.  The resultant
redshifts seem very accurate.  The parameter space proximity and
neural network methods identify 82.6\% and 88.5\% of the lines in the
evaluation set, and for the neural network method, over 98\% of those
identifications are correct.  The spectral resolution of \deimos\
makes both methods nearly perfect at identifying [\oii], but they make
more mistakes in identifying \ha\ (8.6\% and 5.8\% failure rates).
The methods identify [\oiii] and \hb\ lines as one of the two with
6.0\% and 3.6\% failure rates.  Even this ambiguous identification can
give a redshift more precise than present state-of-the-art photo-$z$s.
Remarkably, the neural network method correctly identifies well over
half of the [\oiii] lines without this ambiguity.  The parameter space
proximity method recovers redshifts for 62.7\% of the 640 single-line
galaxies, and the neural network method recovers redshifts for 73.8\%
of the sample.  Overall, the neural network method seems superior in
both accuracy and recovery rate.

Identifying single emission lines is important to future work in
\deep.  The survey contains about 1,000 serendipitously detected
galaxies (``serendips''), many of which display only one emission
line.  The line identification algorithms may be applied to these
objects and double the number of identified single lines.  (One
concern is that the serendips will occupy a region of parameter space
not populated by \deep\ targets in the training set.)  Additionally,
the wealth of data in Field 1, or the EGS, can increase accuracy and
reduce inconclusive identifications with supplemental broadband
measurements or morphological parameters, especially angular sizes,
from high-resolution images.  EGS photo-$z$s calculated without
spectroscopic information can also constrain redshifts enough to
identify single lines.  Furthermore, all four fields also contain more
information than we use in this work.  For example, surface
brightness, angular size, and the galaxy-galaxy correlation function
all contain information about redshift.  \deep\ catalogs these three
observables, which may be implemented in both single line
identification methods.  Finally, although many single-line galaxy
spectra display very faint continuums, cross-correlations with the
continuums of known-redshift galaxies or templates can reveal more
redshift information.  We plan to address these possibilities in
future work.

\acknowledgments We thank the referee for a helpful report.  We also
thank J.~A. Newman for providing statistics on repeat observations,
and we thank C.~M. Pierce for carefully reading a draft of this paper.
We acknowledge National Science Foundation grants AST0507483 and
AST0507428.  ENK is supported by a NSF Graduate Research Fellowship.
Data herein were obtained at the W.~M. Keck Observatory, which is
operated as a scientific partnership among the California Institute of
Technology, the University of California, and NASA.  The Observatory
was made possible by the generous financial support of the W.~M. Keck
Foundation.

\end{document}